\documentclass{jps-cp}
\usepackage{txfonts} 

\usepackage{bbm}
\usepackage{color}

\title{
  Effects of Charge Symmetry Breaking on Form Factors of the Pion and Kaon}

\author{Parada~T.~P.~\textsc{Hutauruk}$^{1}$, Wolfgang~\textsc{Bentz}$^{2,3}$,
  Ian~C.~\textsc{Clo{\"e}t}$^{4}$, and Anthony~\textsc{Thomas}$^{5}$}

\inst{$^{1}$Asia Pacific Center for Theoretical Physics, Pohang, Gyeongbuk 37673, South Korea \\
  $^{2}$Department of Physics, School of Science, Tokai University,
  4-1-1 Kitakaname, Hiratsuka-shi, Kanagawa 259-1292, Japan \\
  $^{3}$Radiation Laboratory, Nishina Center, RIKEN, Wako, Saitama 351-0198,
  Japan\\
  $^{4}$Physics Division, Argonne National Laboratory, Argonne, Illinois 60439, USA \\
  $^{5}$CSSM and ARC Centre of Excellence for Particle Physics at the Terascale, \\ 
  Department of Physics, University of Adelaide, Adelaide SA 5005, Australia}

\email{parada.hutauruk@apctp.org}

\recdate{January 18, 2019}

\abst{Effects of charge symmetry breaking associated with the $u$ and $d$ quark mass
  difference in the elastic form factors of the pion and kaon are presented.
  We use a confining version of the Nambu--Jona-Lasinio model. The pion and kaon are described
  as a dressed quark and antiquark bound states governed by the Bethe-Salpeter equation, and exhibit
  the properties of Goldstone bosons, with the pion mass difference given by $m_{\pi^{+}}^2 - m_{\pi^{0}}^2
  \propto (m_u -m_d)^2$ as demanded by dynamical chiral symmetry breaking. We found significant charge
  symmetry breaking effects for realistic current quark mass ratios ($m_u/m_d \sim 0.5$) in the quark
  electromagnetic form factors of the pion and kaon. We also report
  effects of charge symmetry breaking on the parton distribution functions,
  which are rather smaller than those found in the electromagnetic form
  factors.}

\kword{charge symmetry breaking, elastic form factors,
  parton distribution functions, Nambu--Jona-Lasinio model}

\begin{document}
\maketitle

\section{Introduction} \label{sec:intro}
In quantum chromodynamics (QCD) CSB effects arise from the mass difference
between the $u$ and $d$ quarks, while the difference in the $u$ and $d$ quark
electric charges is the dominant electroweak effect~\cite{MNS90,LPT10}.
Empirically, CSB effects are clearly evident in the proton-neutron mass difference,
and the differing masses between the charged and neutral pion and kaon states,
where for the pion the difference is purely electromagnetic
up to $\mathcal{O}[(m_u - m_d)^2]$ corrections~\cite{GL82}.
CSB effects in hadron masses have been studied
using dynamical lattice simulation of QED+QCD~\cite{Borsanyi13}.
In a different area, CSB is an important background in the extraction of
the strange electromagnetic form factor and parton distribution functions (PDFs)
of the nucleon. As a final example, we note that CSB in the PDFs of the nucleon
is vital to understanding the NuTeV anomaly.

Recent work has also brought the QED contribution to CSB in PDFs under better control~\cite{WTY15}.
Beyond mass differences and effects in low energy nuclear physics,
such as the Nolen-Schiffer anomaly,
the experimental study of CSB effects is challenging. Definitive experiments are certainly needed,
where promising examples include parity-violating deep inelastic scattering (DIS) on the deuteron
and $\pi^+/\pi^-$ production in semi-inclusive DIS from the nucleon,
both of which are planned at Jefferson Lab. In addition, interesting possibilities exist
at an electron-ion collider,
such as charged current reactions, and using pion-induced Drell-Yan
reactions. In this work we report the effect of CSB arising from
the $u$ and $d$ quark mass difference in the leading-twist PDFs and electromagnetic
form factors of the pion and kaon. This report is based on the recent article~\cite{HBCT18}.

\section{Charge symmetry breaking and the NJL model}
%
The NJL model is a quark-level chiral effective field theory of QCD
which shares the same global symmetries as QCD, and is
a Poincar\'e covariant quantum field theory that exhibits dynamical chiral symmetry breaking.
It has been used with success to describe numerous non-perturbative phenomena~\cite{HCT16,Hutauruk18,CBT14}.
The three-flavor NJL Lagrangian, containing only four-fermion interaction terms, has the form
\begin{align}
\mathcal{L}_{NJL} &= \bar{\psi}(i \partial\!\!\!/ 
- \hat{m})\psi  
+ G_{\pi} \left[ (\bar{\psi} \lambda_{a} \psi)^{2} 
- (\bar{\psi} \lambda_{a} \gamma_{5} \psi)^{2} \right]
- G_\rho\left[(\bar{\psi} \lambda_{a} \gamma^\mu \psi)^{2} 
+ (\bar{\psi} \lambda_{a} \gamma^\mu \gamma_{5} \psi)^{2}\right],
\label{eq:njl_lagrangian}
\end{align}
where the quark field has the flavor components $\psi^T = (u, d, s)$,  $\hat{m} = {\rm diag}(m_u, m_d, m_s)$
denotes the current quark mass matrix, $G_{\pi}$ and $G_\rho$ are four-fermion  coupling constants,
and $\lambda_0,\dots,\lambda_8$ are the Gell-Mann  matrices in flavor space
where $\lambda_0 \equiv \sqrt{2/3} \mathbbm{1}$.

The pion and kaon are given as relativistic bound-states of a dressed-quark and
a dressed-antiquark whose properties are determined by solving the $\bar{q}q$
Bethe-Salpeter equation (BSE) in the pseudoscalar channel~\cite{HCT16,CBT14}.
The masses of the pseudocalar mesons are given by the poles $1 + 2 G_\pi \Pi_\alpha (q^2 = m_\alpha^2) =0$,
where $\alpha=\pi^{\pm}, \pi^0, K^{\pm}, K^0, \bar{K}^0$ and the bubble diagrams take the form
\begin{align}
\label{eq:bubblegraphtot}
\Pi_{\alpha}(q^2) &= i\! \int\! \frac{d^4k}{(2\pi)^4}\ 
\mathrm{Tr}\!\left[\gamma_5 \lambda_\alpha^\dagger S(k) \gamma_5 \lambda_\alpha S(k+q) \right],
\end{align}
where the trace is over Dirac, color and flavor indices.
For more details about the NJL models and dressed quark photon vertex as well as the gap equation,
we refer the readers to Ref.~\cite{HBCT18}. Results for the current and dressed quark masses, neutral pion,
and kaon masses, neutral pion leptonic decay constant and meson-quark coupling constants and
the NJL model parameters that vary with $m_u / m_d$ are listed in Table~\ref{tab:parameters}
(See Ref.~\cite{HBCT18} for details).

\begin{table}[tbh]
\caption{Results for the dressed quark masses, neutral pion, and kaon masses,
    neutral pion leptonic decay constant, meson-quark-quark coupling constants, and
    the model parameters that vary with $m_u/m_d$. Recall, that the mass and decay constant of the charged pions
    are fixed at their physical values and therefore do not vary with $m_u/m_d$.
    Similarly, the strange quark mass is kept constant as CSB effects are introduced.
    Note, dimensioned quantities are in units of MeV, with the exception of $G_{\pi}$ which are in units of GeV$^{-2}$.}
\begin{tabular}{ccccccccccccccc}
\hline\hline
$m_u/m_d$  & $M_u$ & $M_d$ & $m_{\pi^0}$ & $m_{K^\pm}$ & $m_{K^0}$ & $f_{\pi^0}$ & $Z_{\pi^0}$ & $Z_{\pi^\pm}$ & $Z_{K^\pm}$ & $Z_{K^0}$ & $G_\pi$ & \\[0.2em]
\hline
0        & 387 & 412 & 137.84 & 483 & 507 & 92.83 & 17.830 & 17.842 & 20.73 & 21.04 & 19.06  \\
0.1       & 390 & 410 & 138.56 & 486 & 504 & 92.89 & 17.837 & 17.846 & 20.76 & 21.01 & 19.05  \\
0.3      & 393 & 406 & 139.38 & 489 & 501 & 92.95 & 17.846 & 17.850 & 20.80 & 20.97 & 19.05  \\
0.5      & 396 & 404 & 139.76 & 491 & 499 & 92.98 & 17.850 & 17.852 & 20.83 & 20.94 & 19.05 \\
0.7      & 398 & 402 & 139.93 & 493 & 497  & 92.99 & 17.852 & 17.853 & 20.86 & 20.91 & 19.04 \\
0.9      & 399 & 401 & 139.99 & 494 & 496  & 93.00 & 17.853 & 17.853 & 20.88 & 20.89 & 19.04\\
1        & 400 & 400 & 140    & 495 & 495 & 93    & 17.853 & 17.853 & 20.89 & 20.89 & 19.04 \\
\hline\hline
\end{tabular}
\label{tab:parameters}
\end{table}
\section{Charge symmetry breaking in pseudoscalar form factors}

\begin{figure}[tbp]
\centering\includegraphics[width=0.65\columnwidth]{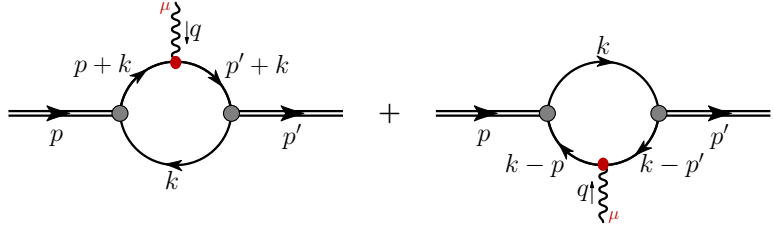}
\caption{Feynman diagrams representing the electromagnetic current of the pion or kaon.} 
\label{fig:emvertex1}
\end{figure}

The matrix element of the electromagnetic current for a pseudoscalar meson $\alpha$ is given by
a single form factor
\begin{eqnarray}
  \mathcal{J}^\mu_\alpha (p',p) &=& (p^{'\mu} + p^\mu ) F_\alpha (Q^2),
\end{eqnarray}
where $p^\mu$ and $p^{'\mu}$ are the initial and final hadron momentum, respectively.
In the NJL model the form factor of a pseudoscalar meson is given by the sum of the two
Feynman diagrams is shown in Fig.~\ref{fig:emvertex1}.
Evaluating these diagrams we find the pseudoscalar form
factors are given by
\begin{align}
\label{eq:fullpi}
F_{\pi^{+}}(Q^2) &= F_{1U}(Q^2) f^{ud}_{\pi^+}(Q^2) - F_{1D}(Q^2) f^{du}_{\pi^+}(Q^2), \\
F_{K^{+}}(Q^2) &= F_{1U}(Q^2) f^{us}_{K^+}(Q^2) - F_{1S}(Q^2) f^{su}_{K^+}(Q^2), \\
\label{eq:fullK0}
F_{K^0}(Q^2) &= F_{1D}(Q^2) f^{ds}_{K^0}(Q^2) - F_{1S}(Q^2) f^{sd}_{K^0}(Q^2),
\end{align}
where the $F_{1Q} (Q^2)$ and $f^{a b}_{\alpha} (Q^2)$ are the dressed charge form factors with $Q=U,D,S$ and
universal body form factors, respectively (See Eqs.~(16) and~(24) of Ref.~\cite{HBCT18} for details).
%

\section{Numerical results}
\begin{figure}[t]
  \centering
  \includegraphics[width=0.45\columnwidth]{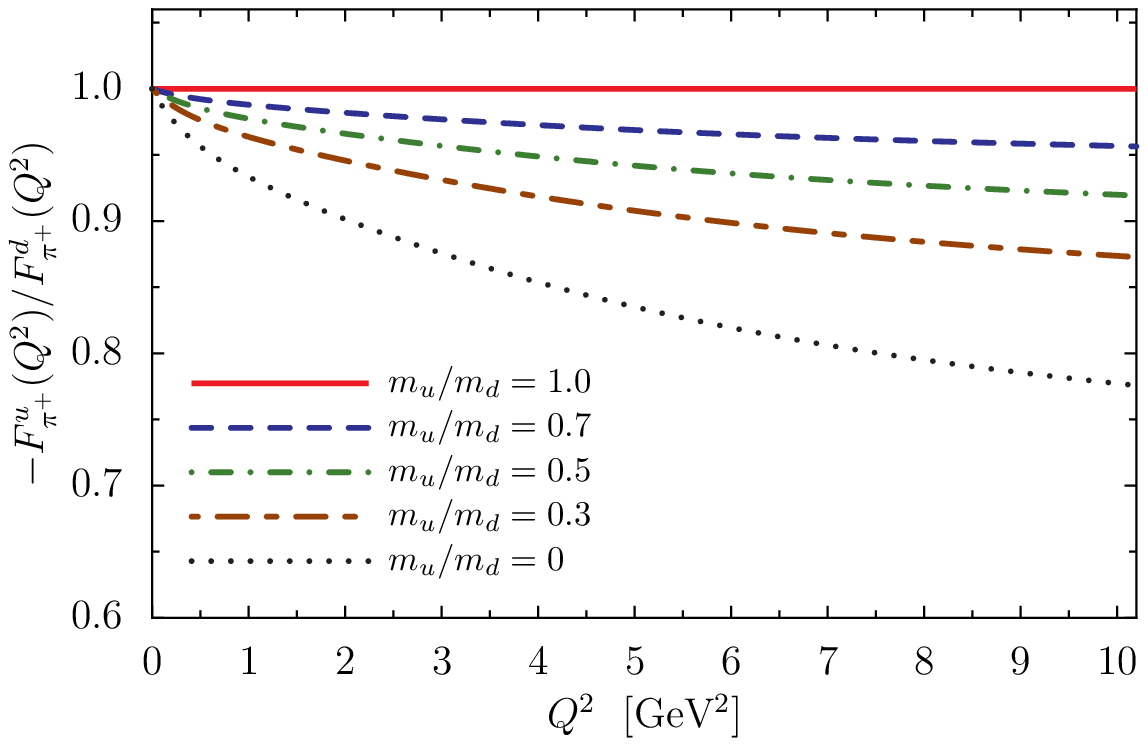} \hspace{0.1cm}
  \includegraphics[width=0.45\columnwidth]{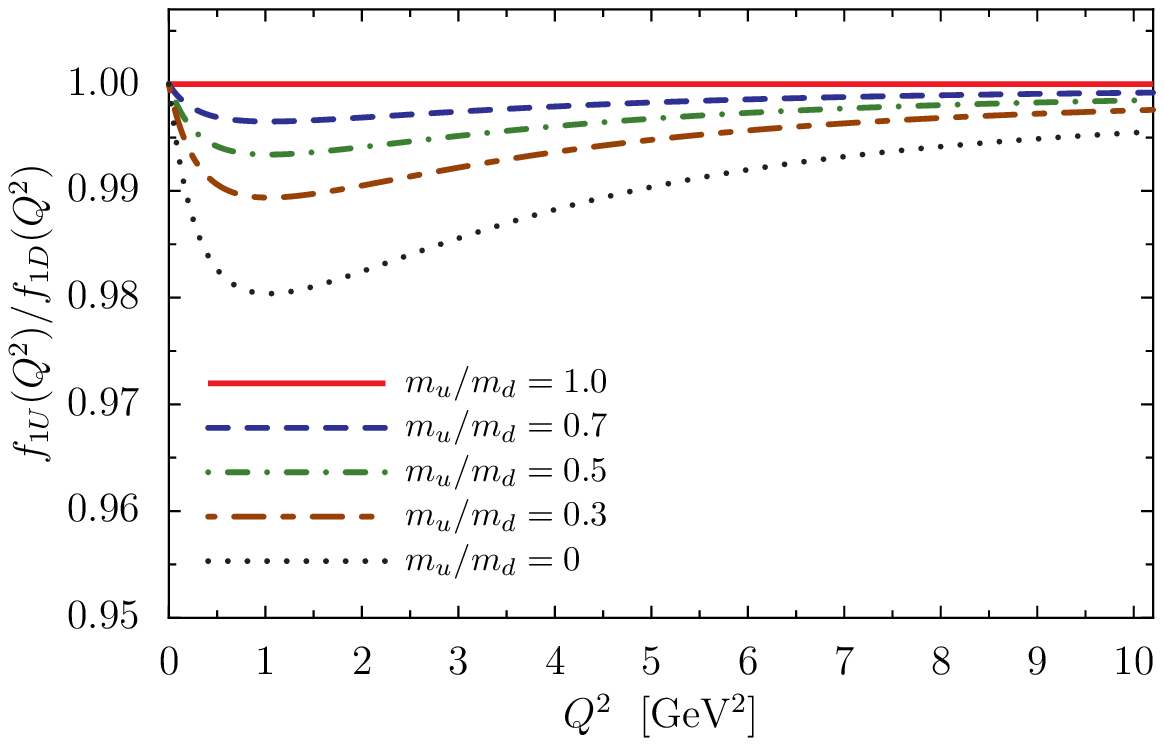}
  \caption{ \label{fig:csvformfactor1}
    Ratio of the $u$ and $d$ quark form factors
    in the $\pi^+$ for various values of $m_u/m_d$ (left panel). CSB effects in the dressed $u$ and $d$
    quark-photon vertex (right panel).}
  \vspace{-2ex}
\end{figure}

Results for the ratio $F^u_{\pi^+} (Q^2)/F^d_{\pi^+}$ at various values of $m_u/m_d$
are shown in the left panel of Fig.~\ref{fig:csvformfactor1}. We
find this ratio decreases from unity as $m_u/m_d$ gets smaller, which reflects that the
$u$ quark charge radius is larger in magnitude than the $d$ quark charge radius.
We find the CSB effects of the size $[|r^u_{\pi^+}|-|r^d_{\pi^+} |]/[|r^u_{\pi^+}|+|r^d_{\pi^+} |] \simeq $
0.7 \%. CSB effects increase substantially with increasing $Q^2$,
reaching 8\% at $Q^2 \simeq$ 10 $GeV^2$ for realistic values of $m_u/m_d$.
This interesting result
is traced to the body form factors, because CSB effects in the quark-photon vertex are small
and vanish for increasing $Q^2$, as depicted in the right panel of Fig.~\ref{fig:csvformfactor1}.

\begin{figure}[t]
  \centering
  \includegraphics[width=0.45\columnwidth]{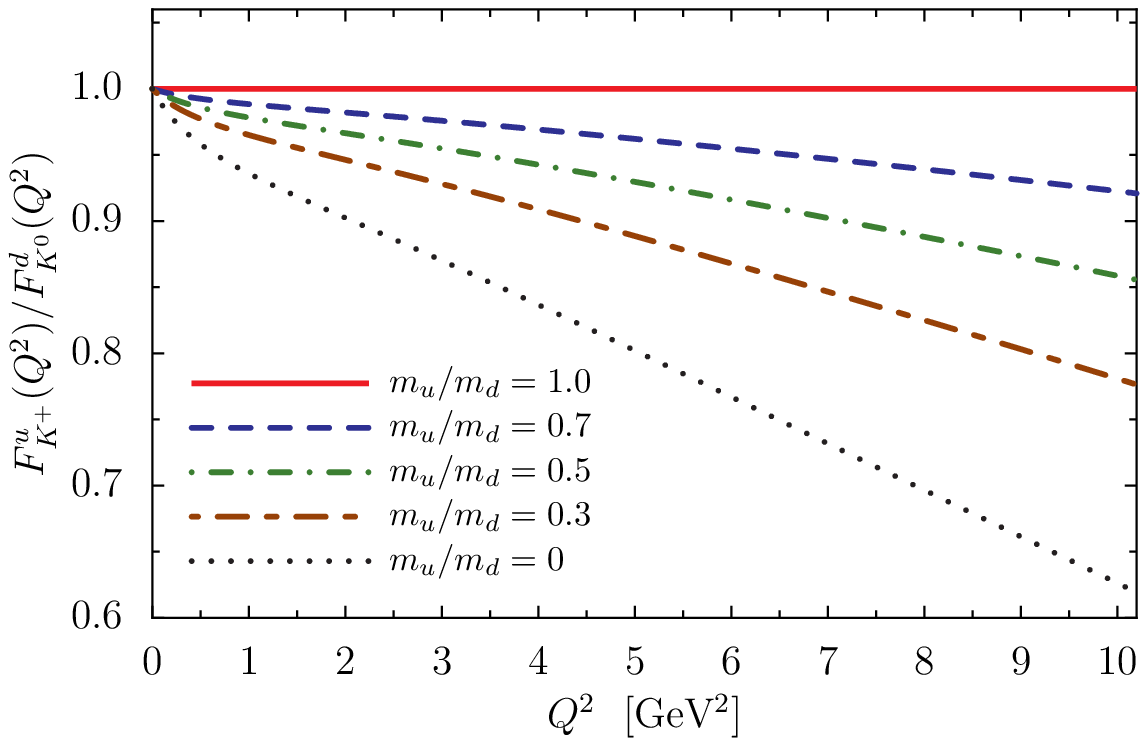} \hspace{0.1cm}
  \includegraphics[width=0.45\columnwidth]{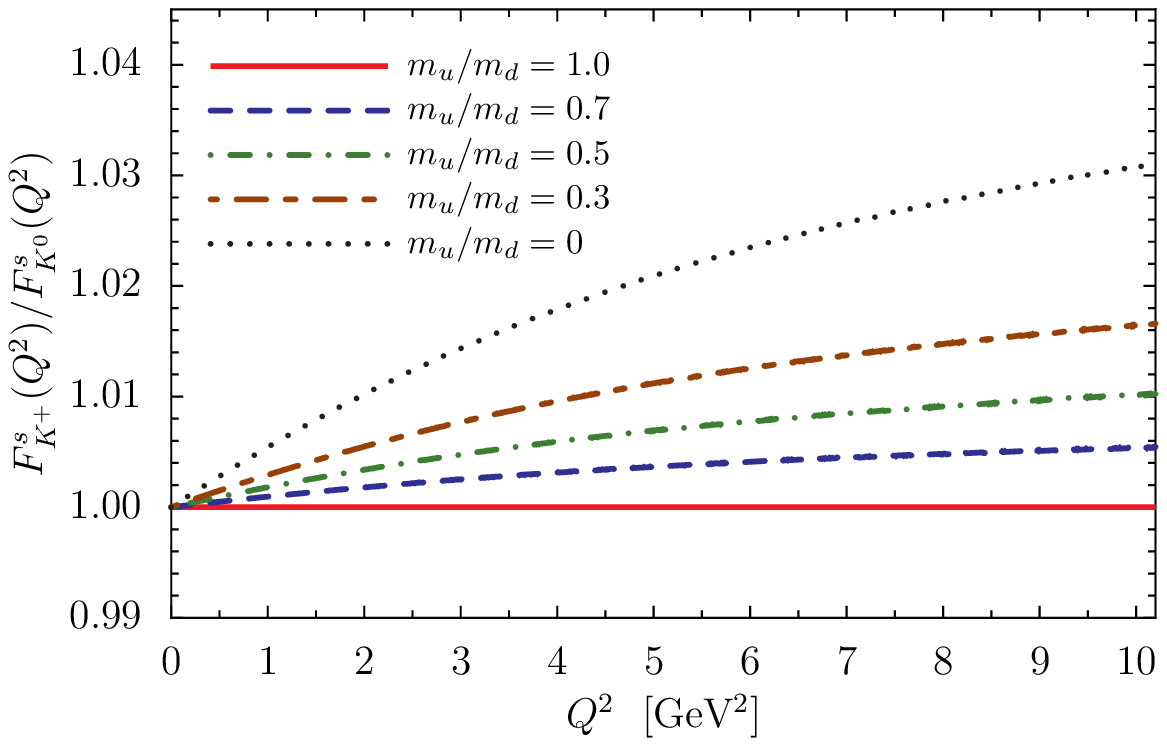}
  \caption{ \label{fig:csvformfactor2}
    Results for CSB effects between
    the $u$ quark sector form factors in the $K^+$ and the $d$ quark sector form factors in the $K^0$ (left panel).
    These CSB effects are found to be about twice that of the pion.
    Comparison between the $s$ quark sectors in $K^+$ and the $K^0$  (right panel).}
   \vspace{-2ex}
\end{figure}

In the left panel of Fig.~\ref{fig:csvformfactor2}, we find that the ratio $F^u_{K^+} (Q^2) / F^d_{K^0} (Q^2)$
is smaller than unity and that the CSB effects grow with increasing $Q^2$.
We therefore find that the $u$ quark charge radius in the $K^+$ is larger in magnitude
than the $d$ quark radius in the $K^0$, which is in agreement with expectation from
the fact that $M_u < M_d$. For $m_u/m_d \simeq 0.5$ we find CSB effects in the quark sector
radii of $[|r^u_{K^+}| - |r^d_{K^0}|]/[|r^u_{K^+}| + |r^d_{K^0}|] \simeq 0.6$\%
which is similar to that found in the pion.

%

The right panel of Fig.~\ref{fig:csvformfactor2} illustrates the ratio $F^s_{K^+}(Q^2)/F^s_{K^0}(Q^2)$
for various values of $m_u/m_d$. We find that this ratio is larger than unity,
which implies that the $s$-quark charge radius in the $K^+$ is smaller
in magnitude than the same radius in the $K^0$ (See Ref.~\cite{HBCT18} for details on the kaon charge radius).
This is consistent with a simple picture for the kaon, where the lighter $u$ quark
is less able to pull the heavier $s$ quark away from the charge center of kaon.
We note however, that these environment sensitivity effects are at the few percent level,
and therefore much smaller than the CSB effects. Additionally,
we report the CSB effects in the parton distribution functions
(See Ref.~\cite{HBCT18} in more details). We find that the CSB effects in the PDFs are much
smaller than in the electromagnetic form factors at high momentum transfer $Q^2$. At
scale of $Q^2 =$ 5 GeV$^2$, we find that the CSB effects are at few percent level,
showing CSB effects in the kaon PDFs much smaller than these effects in the pion.

\section{Summary}
To summarize, we have revisited CSB effects in the spacelike electromagnetic form factors
of the pion and kaon using the NJL model
with the proper-time regularization scheme.

We found that the effect of CSB arising
from the light quark mass differences is surprisingly large in the quark
elastic form factors at large momentum transfer. This is especially dramatic in the kaon,
where for a realistic value of $m_u/m_d \simeq 0.5$ one finds CSB at the 15\% level
in the ratio $F^u_{K^+}(Q^2)/F^d_{K^0}(Q^2)$ at $Q^2 \simeq 10\,$GeV$^2$.
The analogous changes in the quark distribution functions are considerably smaller in magnitude,
reaching 3\% as $x\rightarrow 1$ in the pion ratio $u_{\pi^+}(x)/\bar{d}_{\pi^+}(x)$, compared
with just 1\% in the ratio $u_{K^+}(x)/d_{K^0}(x)$ for the kaon.
It would also be of interest to explore the expected degree of CSB in these systems
using other realistic models and lattice QCD.

\end{document}